\documentclass[twocolumn,showpacs,preprintnumbers,superscriptaddress,prl]{revtex4-1}
\usepackage{graphicx}
\usepackage{dcolumn}
\usepackage{bm}
\usepackage{times}
\usepackage{natbib}
\usepackage{color}
\definecolor{Blue}{rgb}{0.3,0.3,0.9}
\definecolor{Red}{rgb}{0.9,0.3,0.3}
\definecolor{Green}{rgb}{0.3,0.9,0.3}

\begin{document}

\title{Evidence of deep water penetration in silica during stress corrosion fracture}

\author{F. Lechenault}
\affiliation{ CEA-Saclay, IRAMIS, SPEC, F-91191 Gif-sur-Yvette, France}
\author{C. L. Rountree}
\affiliation{ CEA-Saclay, IRAMIS, SPCSI, F-91191 Gif-sur-Yvette, France}
\author{F. Cousin}
\affiliation{CEA-Saclay, IRAMIS, Laboratoire L\'eon Brillouin,  F-91191 Gif-sur-Yvette, France}
\author{J.-P. Bouchaud}
\affiliation{Science and Finance, Capital Fund Management, 6 Boulevard Haussmann, 75009 Paris, France}
\author{L. Ponson}
\affiliation{Graduate Aerospace Laboratories (GALCIT), California
Institute of Technology, Pasadena, CA 91125, USA}
\author{E. Bouchaud}
\affiliation{ CEA-Saclay, IRAMIS, SPEC, F-91191 Gif-sur-Yvette, France}
\affiliation{Norwegian University of Science and Technology, Department of Physics, Hogskoleringen 5, N-7491 Trondheim, Norway }

\date{\today}

\begin{abstract}

We measure the thickness of the heavy water layer trapped under the stress corrosion fracture surface of silica using neutron reflectivity experiments. 
We show that the penetration depth is 65-85 \aa ngstr\"{o}ms, suggesting the presence of a damaged zone of $\sim $100 \aa ngstr\"{o}ms extending ahead of the crack tip during its propagation.
This estimate of the size of the damaged zone is compatible with other recent results. 

\end{abstract}


\maketitle

If a small stone hits your windshield and you see cracks growing slowly from the impact, you are probably observing a stress corrosion process. Under very moderate external tensile stresses, cracks may indeed grow in silicate glasses, thanks to a chemical reaction which involves the water molecules of the surrounding environment. This is a complex phenomenon, which started to be studied  in the sixties~\cite{Wieder68_IntJFractMech,Wieder69_JAmCerSoc} and is not yet fully understood (see~\cite{Ciccotti09_jpd} for a recent review). As a matter of fact, two important questions remain to be solved: the exact mechanism by which water molecules manage to break Si-O bonds, and the role of the amorphous structure of glass in the fracture properties. 

In the classical picture, proposed first by Michalske and Bunker~\cite{Michalske84_JApplPhys}, water molecules break the Si-O bonds located exactly at the crack tip thanks to a hydrolysis reaction. For small enough external loads, the crack velocity is controlled by the rate of the chemical reaction, which depends both on the degree of ambient humidity and on the applied stress. This regime is traditionally refered to as ``Stage I''~\cite{Wieder68_IntJFractMech,Lawn93_Book}. At higher applied loads, when the crack velocity reaches a value that depends on humidity, the slowest phenomenon (which imposes its kinetics to the crack velocity) is the diffusion of water molecules to the tip along the fracture surfaces. Since surface diffusion is not sensitive to the external applied load, the crack velocity in this ``Stage II'' does not depend on it either. 

This classical picture, however, does not take into account the disordered character of the glass structure. In a perfect crystal, where atomic bond orientations and energies are $\delta$-distributed, 
bonds at the crack tip will break first because stress concentration is maximum there. But when considering a more complex arrangement of chemical bonds, disjunctions are likely to occur at some distance 
away from the tip. This effect was seen first in Molecular Dynamics (MD) simulations of dynamic fracture, where a nanometric Process Zone (PZ) was observed to form ahead of the main crack front~\cite{Nakano95_prl,Rountree02_anrevmatres,Chen07_prl}. Somewhat later, such a ``quasi-brittle''~\cite{Lawn93_Book,Bonamy10_pr} behavior was claimed to be observed experimentally using {\it in situ} Atomic Force Microscopy (AFM) experiments~\cite{Celarie03_prl,Prades04_IntJSolStruct, Rountree10_PhysChemGlass}. These observations are 
however still very controversial~\cite{Guin04_prl,Lopez-Ceperro07_IntJMaterRes}. Because AFM observations are restricted to the free surface of the specimens, several artifacts can alter the measurements~\cite{Fett08_prb}. Moreover, results may be tampered by tip effects~\cite{Lechenault_prl10}. As a matter of fact, there are several significant differences between the free surface and the bulk of the specimen. Of particular importance is the exposure to water in the case of stress corrosion fracture: while the free surface is in direct contact with the ambient humidity, water molecules have to diffuse within the material for Si-O bonds to break at a distance from the crack tip. Although experiments have been performed at high temperature only~\cite{Davis96_jncs,Berger03-JNCS}, a rough extrapolation of Tomozawa et al's results~\cite{Tomozawa99_MatSciEngA} suggests a water diffusion coefficient in silica of the order of $\sim $10$^{-21}$cm$^2$.s$^{-1}$ at room temperature. This means that the penetration length of water molecules into unstrained glass should be approximately 3pm  (respectively 0.3\AA) during the time it takes for a crack moving at 10$^{-6}$m.s$^{-1}$ (respectively at 10$^{-8}$m.s$^{-1}$) to cover 100$\mu $m. 

However, because of the huge stresses concentrated at the crack tip, diffusion is enhanced by orders of magnitude in the vicinity of the tip during fracture~\cite{Larche96_defdiffo,Mehrer96_defdiffo}, as observed in several other materials~\cite{Aziz06_PRB,Guery09_PRE}. 
Therefore, water is expected to penetrate into the glass and, because of the heteogeneity of the material mentioned above, start breaking bonds and create microcracks ahead of the crack tip. This 
in turn increases further the diffusion of water, thereby creating more corrosion and potentially leading to a substantial damaged zone. If this scenario is correct, a rather thick layer of water should remain
trapped underneath the nominal fracture surface after the crack has propagated and stresses have relaxed. Since the diffusion constant is so small  in unloaded silica glass (more than 100 days for travelling 1 nm), one
should observe post-mortem a ``fossil'' water profile, essentially frozen-in at the time of its creation, with a thickness of the order of the size of the damaged zone. The aim of this work is to provide quantitative evidence for the above scenario using neutron reflectivity~\cite{Bouchaud86_epl,Cousin_cnrs} to measure the thickness of the water layer left behind the crack. We find that the penetration depth of water is 
of the order of a hundred \aa ngstr\"{o}ms, suggesting the presence of a large damaged zone.

\begin{figure}
\includegraphics[width=0.3\textwidth]{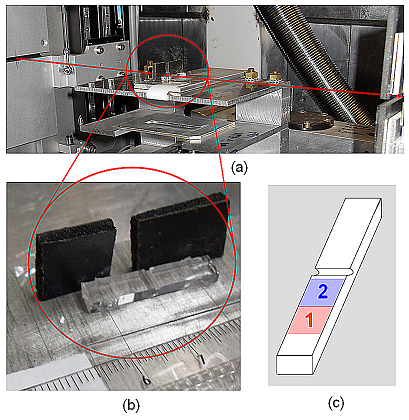}
\caption {(a) Picture of the experimental setup. The neutron beam is schematized in red, incoming from the last slit of the collimator, reflecting on the sample and going within the slit in front of the detector. (b) Picture of the broken sample, showing the two black sheets of B$_4$C used to select one area of interest (see text). (c) Sketch of the broken sample, with the two parts described in the text.}
\label{fig:setup}
\end{figure}

{\it Fracture experiments}. Fracture experiments were conducted in a highly controlled manner via Double Cleavage Drilled Compression (DCDC) samples. In this geometry, the stress at the crack tip naturally decreases, enabling us to conduct all our experiments in the stress corrosion regime.  DCDC samples used herein are cuboids of size 5 x 5 x 25mm$^3$ with a 1mm diameter hole drilled in the center.  They are made of Corning 7980 pure silica.  The fracture experiments were conducted in a glove box which had been saturated with heavy water.  Two symmetrical precracks are first initiated from the hole of the sample as described in~\cite{Prades04_IntJSolStruct}. Subsequently the load is adjusted in order to reach a desired velocity~\cite{Prades04_IntJSolStruct}. Zone 1 (Fig.~\ref{fig:setup}c) corresponds to a stress intensity factor 0.61MPa.m$^{1/2}$.  Zone 2 (Fig.~\ref{fig:setup}c) corresponds to a stress intensity factor 0.77MPa.m$^{1/2}$ (average velocity of 4.10$^{-6}$m.s$^{-1}$).  

Heavy water has been chosen because its coherent length density $b_w=6.39\, 10^{-6}$\AA$^{-2}$ is higher than the one of silica ($b_s= 3.41\, 10^{-6}$\AA$^{-2}$), and of opposite sign to that of light water 
($- 0.53\, 10^{-6}$\AA$^{-2}$). If some water is trapped in the vicinity of the surface of the sample, the reflectivity of the sample should thus increase in the presence of heavy water whereas it would
only weakly decrease with light water.  

{\it Neutron reflectivity experiments}. Specular Neutron Reflectivity (SNR) measurements have been carried out on the horizontal time-of-flight EROS reflectometer (Saclay, France) with a fixed angle $\theta$ of $1.195°$, with a neutron white beam covering wavelengths $\lambda$ from 4\AA ~to 25\AA , covering an accessible $q$-range (diffusion vector $q = 2\pi \sin {\theta }/\lambda $) from 0.005 \AA$^{-1}$ to 0.032\AA$^{-1}$.
 
Zones 1 and 2 (see Fig.~\ref{fig:setup}c)) described above were studied. In order to select one of these areas of interest, we used the following trick. The sample was almost completely hidden on the neutrons path by two black sheets of B$_4$C, a strong neutron absorber, to let the neutrons illuminate only the desired region (Fig.~\ref{fig:setup} a and b). In order to test that the selected region was flat enough to allow a correct measurement, we have checked that the half full width of the alignment rocking curve was lower than $0.25^{\circ}$. When this was not the case, the illuminated region was reduced until this condition was met. The resulting illuminated surfaces were very small, of the order of $\sim 25$mm$^2$. Because of this smallness, we used very long counting times to get a reasonable noise-to-signal ratio, up to 48 hours per illuminated region. In particular, we measured the background independently from the sample with great precision, enabling its subtraction with a good accuracy.  

\begin{figure}
\includegraphics[width=0.5\textwidth]{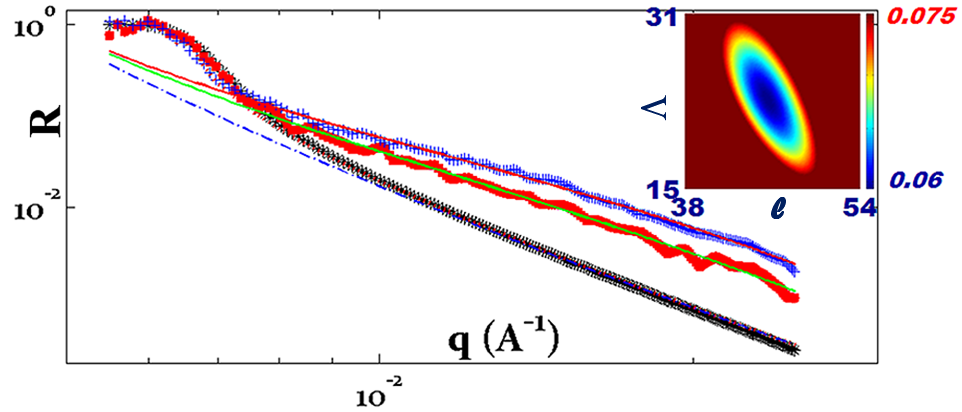}
\caption{Experimental neutron reflectivities plotted as a function of the diffusion vector $q$ for: the control sample ($\star $), the specimen broken at a stress intensity factor $K_I=0.61$MPa.m$^{1/2}$ (Zone 1, {\color{Red} $\star $} ) and for the specimen broken at a stress intensity factor $K_I=0.77$MPa.m$^{1/2}$ (Zone 2, {\color{Blue} +} ). We have superimposed to the experimental curve corresponding to the control sample the theoretical Fresnel reflectivity curve (red dotted line) as well as the result of the second order Born approximation (blue dashed dotted line). The experimental results corresponding to Zone 1 and Zone 2 are fitted using  Eqs.~(\ref{Rlimite}),~(\ref{r1-1}) and~(\ref{r2}). The best fits correspond to $\phi_0^I=0.348 \pm 0.003$, $\ell^I \approx 43$\AA~ and $\Lambda^I \approx 35$\AA~in Zone 1, and  $\phi_0^{II}=0.567 \pm 0.003$, $\ell^{II}\approx 46$\AA~and $\Lambda^{II} \approx 23$\AA~in Zone 2. Inset: Contour lines of the fit root-mean square error in the plane ($\ell, \Lambda$) in Zone 2, showing that while the combination $\ell_{eff}=\Lambda + \ell$ is rather well pinned down by the fit, $\Lambda - \ell$ is a ``soft'' direction. The relative experimental rms error per point is $0.075$, whereas
the minimum relative error achieved by the fit is $0.06$.}
\label{fig:R1}
\end{figure}

{\it Results}. The experimental curves presented in Fig.~\ref{fig:R1} clearly show a huge change in the reflectivity of broken samples when compared to the reflectivity of an unbroken control specimen. Since the roughness of the fracture surfaces is larger than the roughness of the control specimen, one should {\it a priori} expect a decrease of the reflectivity of the broken samples. The difference seen in Fig.~\ref{fig:R1} is therefore underestimated. 

Fig.~\ref{fig:R1} shows also that the reflectivity of the control specimen corresponds perfectly to the Fresnel reflectivity $R_F=(q-\sqrt{(q^2-q_c^2)})^2/(q+\sqrt {(q^2-q_c^2)})^2$ of a semi-infinite silica diopter for which the coherent length density is equal to $b_s$~\cite{Cousin_cnrs,sfn_website}, for which $q_c=\sqrt{4\pi b_s}$. In order to fit the neutron reflectivities measured in Zones 1 and 2 (Fig.~\ref{fig:R1}), we have 
used a second order Born approximation, assuming that the heavy water concentration is not a constant, but that it decreases with the distance $z$ from the free surface of the tested sample as $\phi (z )$. This concentration profile translates into a coherent length density profile $b_w(z)=b_w \phi (z)$. Since most of the workable signal is obtained in a region of large diffusion vectors $q$, far enough from total reflexion, reflectivities are quite small, and hence the reference situation is the free case, when all the incident neutrons are transmitted.  The neutron wave function $\psi $ obeys the following eigenvalue equation: $d^2\psi /dz^2=-q^2\psi + V(z) \psi $, with $V(z)=4\pi (b_s+b_w \phi (z))$ a small perturbation: $V(z) \ll q^2$. We write: $\psi =\psi _0+ \psi _1+ \psi _2+... $, with $\psi _1$ and $\psi _2$ the first and second order corrections (in $V$). Adapting the calculation of~\cite{Charmet83_JOptSocAm}, one finds: $\psi_{n+1}(z) = \int_0^\infty dz' V(z') \psi _n(z')G_0(z,z')$, where the Green function $G_0(z,z'>0)$ relevant for our boundary conditions is: 
\begin{eqnarray}
\begin{array}{ll}
  	G_0(z,z')=-{1\over 2iq}\exp {(-iq(z'-z))}
 	& \mbox{if}\, z<0  \\
	G_0(z,z')=- {1\over 2iq}\exp {(-iq|z'-z|)}
 	& \mbox{if}\, z>0  \label{GreenIII}
  \end{array}
\end{eqnarray}
Using the above expressions, we get the expression of the reflectance $r$ to order $V^2$:
\begin{eqnarray}
& r & =  {i\over 2q}\int_0^{\infty }dz' V(z')e^{-2iqz'}-{1\over 4q^2}\int_0^{\infty }dz' V(z') \times \nonumber \\ 
&\times & \left[e^{-2iqz'}\int_0^{z'}dz''V(z'')+\int_{z'}^{\infty }dz''V(z'')e^{-2iqz''}\right] 
\label{reflectance}
\end{eqnarray}
The reflectivity $R$ is $R=|r|^2$. In order to check the validity of our second order Born approximation, we first verify that our result in Eq.~(\ref{reflectance}) tends to the Fresnel reflectivity for high values of 
$q$ when $V$ is a constant equal to  $q_c^2=4\pi b_s$. This limit leads to the following reflectivity:
\begin{equation}
R=r_0^2={q_c^4\over 16 q^4}\left(1+{q_c^2\over 2q^2}\right)^2+O({q_c^8 \over q^8})
\label{Rlimite}
\end{equation}
which coincides with the corresponding large $q$ expansion of the Fresnel reflectivity (see Fig.~\ref{fig:R1}).

We then tried to fit the reflectivities in Zones 1 and 2 using the simplest function involving a single length scale, i.e. $\phi (z)=\phi_0 \exp {(-z/\Lambda )}$. Although this can be made to fit the Zone 1 results, the reflectivity increase in Zone 2 is too large to be accounted for using this simple function. Hence, guided by the idea that there might be a saturated layer of depth 
$\ell$ close to the surface that becomes more diffuse deeper in the sample, we posit that: 
\begin{eqnarray}
\begin{array}{ll}
  	\phi _w(z)=\phi _0
 	& \mbox{if } z<\ell\\
	\phi _w(z)=\phi _0 \exp {(-(z-\ell)/\Lambda ) }
 	& \mbox{if } z>\ell \label{newprof}
  \end{array}
\end{eqnarray}
This choice leads to a reflectance $r$ that can be written as: $r=r_0+r_1+r_2$, with $r_0$ as the Fresnel reflectance (Eq.~(\ref{Rlimite})) and:
\begin{equation}
r_1  = - {2\pi b_w\phi _0\over q(2iq\Lambda +1)} \exp {(-2iq\ell)}
\label{r1-1}
\end{equation}

\begin{eqnarray}
r_2 & = & -{4 \pi^2(b_w\phi _0)^2\over q^2}\left[{1\over 2q^2}[\exp {(-2iq\ell)}(1+2iq\ell)-1]\right.\nonumber \\  &+ & \left.{\Lambda \exp {(-2iq\ell) }\over (2iq\Lambda +1)(iq\Lambda +1)}
(\Lambda +2\ell+2iq\Lambda \ell)\right]
\label{r2}
\end{eqnarray}

Fig.~\ref{fig:R1}  shows the best fits of the experimental measurements performed on the two fracture surfaces using $R=|r_0+r_1+r_2|^2$ (Eqs.~(\ref{Rlimite}),~(\ref{r1-1}) and~(\ref{r2})). 
The best fit is achieved with $\phi_0^I=0.348 \pm 0.003$, $\ell^I \approx 43$\AA~ and $\Lambda^I \approx 35$\AA~in (slow) Zone 1, while in (fast) Zone 2, $\phi_0^{II}=0.567 \pm 0.003$, $\ell^{II}\approx 46$\AA~and $\Lambda^{II} \approx 23$\AA. Note that $\phi_0$ is very accurately determined by the fit, although the error bar we quote only accounts for statistical uncertainty, and not systematic effects 
coming from the choice of the fitting function and of the interval over which the data is fitted. On the other hand, the quality of fit has a ``soft direction'' in the plane ($\ell,\Lambda$), as represented in the inset of Fig.~\ref{fig:R1}. As expected, the total effective width of the layer, $\ell_{eff}=\ell+\Lambda$ is better determined than $\ell$ and $\Lambda$ separately. The statistical error bar on $\ell_{eff}$ is smaller than $1$\AA, but again systematic errors are much larger. As shown in the inset of Fig.~\ref{fig:R1}, $\ell_{eff}$ can be varied by $\sim \pm 5$\AA~and still lead to an acceptable fit. 

{\it Discussion}. Our results therefore clearly show that heavy water is present over $\sim 65$ to $85$\AA~ under the stress corrosion fracture surface of pure silica in the conditions of our experiments. This penetration depth is much larger than what is expected using a room temperature extrapolation of the diffusion coefficient of light water in silica~\cite{Tomozawa99_MatSciEngA} (to our knowledge, there are no such results concerning heavy water, for which diffusion should be even smaller). A likely explanation is as follows: diffusion enhancement in the vicinity of the crack tip, where huge tensile stresses are present~(\cite{Larche96_defdiffo} -~\cite{Guery09_PRE}), allows the water to penetrate inside the bulk and create a damaged zone which helps water progressing further still. Our neutron scattering experiment gives information about the water penetration depth in the direction perpendicular to the fracture surface. To investigate how far water penetrates parallel to the crack propagation direction, one would need a detailed self-consistent model for the coupled growth of the damaged zone and the diffusion of water. We grossly simplify the problem by computing the dilation field $D(x,y) = \epsilon_{xx}(x,y) + \epsilon_{yy}(x,y)$ in the vicinity of the crack tip within a purely elastic model, and postulating that the water penetrates in a region 
defined by $D(x,y) > D_c$, and infer the anisotropy of the water penetration from that of the iso-dilation lines. We analyze DCDC samples identical to those used in the experiments and compute the dilation field in the mid-plane of the specimen using finite element simulations, with element sizes decreasing exponentially as approaching the crack tip, so that the strain field is resolved at the nanometer scale within that region. As shown in Fig. 3 for an external loading $K = 0.61$ MPa.m$^{1/2}$, the domain at the crack tip with a high level of dilation extends deeper in the direction of propagation than in the perpedicular direction, by a factor $\sqrt{2}$. Therefore, we estimate that water penetrates 
roughly~$9-12$nm ahead of the crack tip, in the crack propagation direction. If one identifies the water-rich region with a damaged zone, our estimate is in agreement with the fact that the strain field observed in the vicinity of a stress corrosion crack tip is elastic only on scales larger than $\sim 10$nm~\cite{Pallares10_these}.

\begin{figure}
\includegraphics[width=0.4\textwidth]{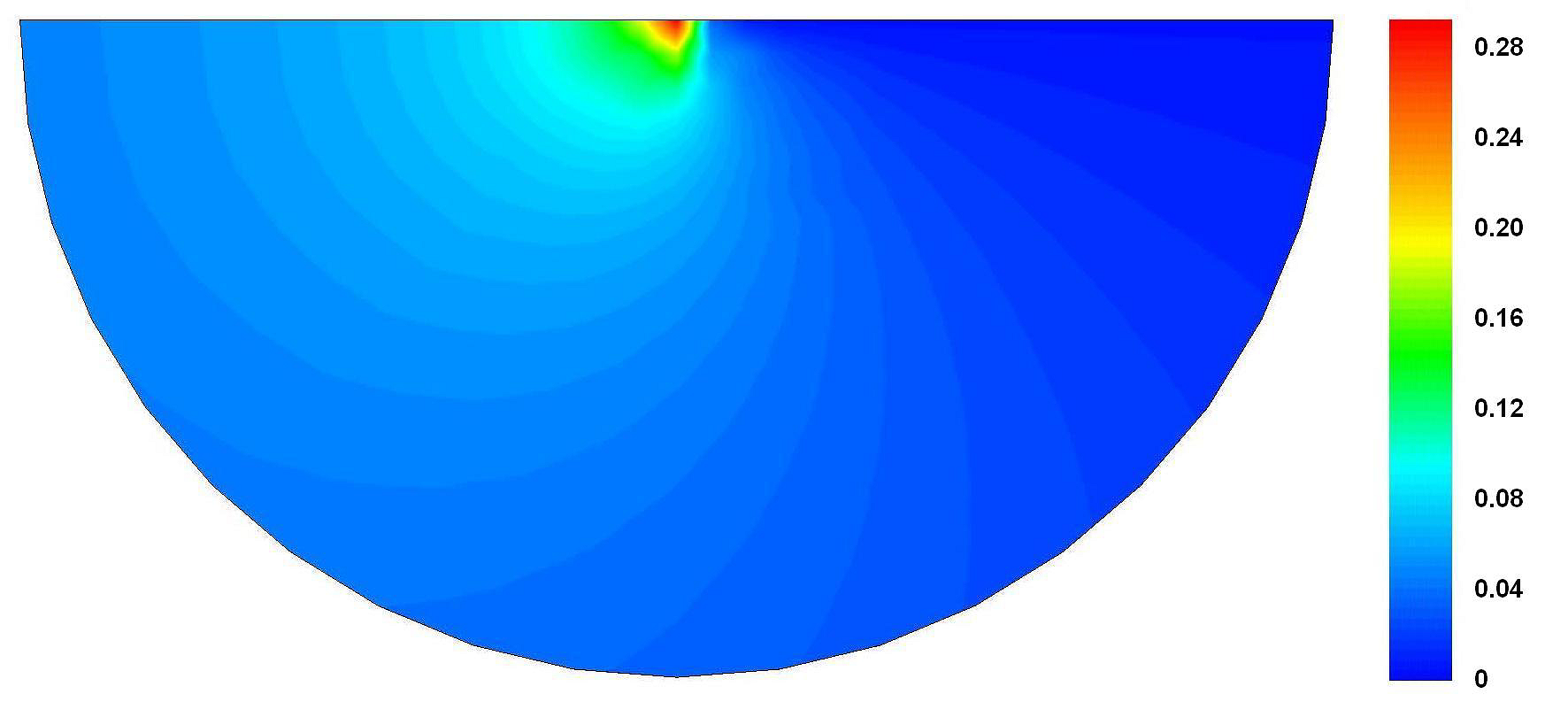}
\caption{Finite element simulations of the dilation field $D(x,y) = \epsilon_{xx}(x,y) + \epsilon_{yy}(x,y)$ in the vicinity of the crack tip in the mid-plane of the specimen for an external loading $K = 0.61$ MPa.m$^{1/2}$.
The crack propagates from right to left, and the color code indicates the amplitude of $D(x,y)$.}
\label{fig:FE}
\end{figure}

We observe an increase of the neutron reflectivity, not only with respect to the Fresnel case, but also with respect to the case where we assume a heavy water single-length scale, exponentially decreasing profile. This has prompted us to postulate the presence of an homogeneous layer of water, followed by
an exponentially decaying profile. The width of the layer, $\ell \approx 4$nm, is significantly larger than the height fluctuations on silica fracture surfaces which do not exceed $1$nm~\cite{Salaud06_prl,Ponson07_annales}. It is tempting to interpret this zone as a strongly damaged zone, with a density of microcracks that is larger when  the stress intensity factor -- and the crack velocity -- is higher. This would naturally explain why 
$\phi_0^{II} > \phi_0^I$, although we have no clear explanation as why $\ell^{I} \approx \ell^{II}$. This might be due to the cancellation of 
two opposite effects: a stronger stress enhances the diffusion of water in the bulk, but at the same time the crack speed is larger, leaving less 
time for the corrosion mechanism to operate. The latter mechanism in fact explains why the exponential region, which is probably more 
sensitive to diffusion, is wider in Zone 1 than in Zone 2 ($\Lambda^{I} > \Lambda^{II}$). In order to be more quantitative, one requires at this 
stage a detailed model for the formation of the damaged zone, dynamically coupled to the water profile.  
A more systematic study of the effect of the external applied stress on the heavy water content would be needed to provide a sound basis for such a quantitative model. Because the damaged zone size is expected to be even larger for lower crack velocities~\cite{Salaud06_prl,Ponson07_annales}, further experiments will be performed for crack speeds as low as $10^{-11}-10^{-10}$m.s$^{-1}$, with larger specimens such as to increase the intensity of the reflected neutron beam.

{\bf Acknowledgements:} Part of this work has been done within the CORCOSIL ANR project. The authors thank M. Ciccotti, G. Pallares, K. Ravi-Chandar and M. Tomozawa for very interesting discussions.

\bibliographystyle{unsrt}
\bibliography{bibneutronsJ}

\end{document}